\documentclass[preprint,12pt]{elsarticle}

\usepackage{tabularx}
\usepackage{mathptmx}
\usepackage{etoolbox}
\usepackage{amsmath}
\usepackage{braket}
\usepackage{dsfont}
\usepackage{units}
\usepackage{placeins}
\usepackage{comment}
\usepackage{nicematrix}
\usepackage{caption}
\usepackage{booktabs}
\usepackage{parnotes}
\usepackage{textgreek} 
\journal{Physics Open}

\begin{document}

\begin{frontmatter}

\title{Design of a Lamb-Shift Polarimeter\\
for $^3$He Ions and Atoms} 
\author[first_affiliation,second_affiliation,third_affiliation]{N.~Faatz} 

\cortext[cor1]{Corresponding author} 
\affiliation[first_affiliation]
{organization={GSI, Helmholtzzentrum~für~Schwerionenforschung},\\
                addressline={Planckstraße~1,},
                 city={Darmstadt},
                 postcode={64291},
                 country={Germany}}


\affiliation[second_affiliation]
  {organization={Institut für Kernphysik, Forschungszentrum~Jülich},\\
                addressline={Wilhelm-Johnen-Straße~1,},
                 city={Jülich},
                 postcode={52428},
                 country={Germany}}               
\affiliation[third_affiliation]{organization={III. Physikalisches Institut B, RWTH Aachen},
                addressline={Templergraben~55},
                 city={Aachen},
                 postcode={52062},
                 country={Germany}}

\affiliation[fourth_affiliation]{organization={Heinrich-Heine-Universität Düsseldorf},\\
                addressline={Universitätsstraße~1,},
                 city={Düsseldorf},
                 postcode={40225},
                 country={Germany}}

\affiliation[fifth_affiliation]{organization={Institut für Kernphysik, Universität zu Köln},
                addressline={Zülpicher Straße~77},
                city={Köln},
                postcode={50937},
                country={Germany}}

\affiliation[sixth_affiliation]{organization={FH Aachen Campus Jülich},
                addressline={Heinrich-Mußmann-Straße~1},
                city={Jülich},
                postcode={52428},
                country={Germany}}

\affiliation[seventh_affiliation]{organization={Peter-Grünberg-Institut, Forschungszentrum~Jülich},\\
                addressline={Wilhelm-Johnen-Straße~1,},
                 city={Jülich},
                 postcode={52428},
                 country={Germany}}

\author[first_affiliation,second_affiliation]{R.~Engels\corref{cor1}}
\ead{r.engels@gsi.de}
\author[fourth_affiliation]{C.~Kannis}
\author[first_affiliation,second_affiliation,fifth_affiliation]{S.~J.~Pütz}
\author[sixth_affiliation,seventh_affiliation]{J.~Steinhage}



\begin{abstract}
In high energy physics experiments $^3\text{He}^{2+}$ nucleons are suitable to test the internal structure of neutrons, which makes them the perfect substitute. Moreover, it is one of the atomic species for which high polarization values are achievable through the use of optical pumping methods. Therefore, nuclear polarized sources for $^3\text{He}^{2+}$ ions are currently in development. Consequently, detector systems validating their nuclear polarization value need to be established. Such a detector system, based on the Lamb-shift polarimeter, is introduced in this work. Its advantage is that it can operate at low energies in the range of 10 to 100 keV, which makes pre-acceleration unnecessary. In addition, a fast evaluation of the nuclear polarization within seconds leads to shorter disruptions for beam times.
Another option of this Lamb-shift polarimeter might be that it can serve as a valuable tool for further investigation of the bound tritium decay into $^3$He atoms, similar to the hydrogen spin filter for the bound beta decay of the free neutron.
\end{abstract}

\begin{keyword}
Nuclear spin filter, Nuclear spin polarization, Lamb-shift polarimeter, metastable helium-3
\end{keyword}

\end{frontmatter}


\newpage
\section{Introduction}
The $^3\text{He}$ atom is one of the stable isotopes of helium and plays an important role in the world of physics. Due to the unpaired neutron in its nucleus, its magnetic moment is similar to that of a free neutron~\cite{RHIC_neutron_sub_exp,RHIC_neutron_sub_theory}. Consequently, it enables research of the neutron structure. As an ion, the trajectories of $^3\text{He}^{2+}$ can be manipulated and, therefore, the ions can be stored in storage rings, which increases the luminosity of many experiments. 
\newline
In particular, experiments with nuclear spin polarization are of great interest. In the 20th century first polarized $^3\text{He}^{2+}$ ion sources have been built based on the Stern-Gerlach~\cite{Helium_ABS_Quelle}, Lamb-shift~\cite{Helium_Lamb_Quelle} and optical pumping~\cite{Helium_pumping_source} methods. In the meantime, Maxwell et al.~are working on realizing a new nuclear spin polarized $^3\text{He}^{2+}$ source based on optical-pumping methods~\cite{RHIC_3He_source}, which will be implemented in the future electron-ion collider (EIC) at Brookhaven National Laboratory (BNL), once completed~\cite{RHIC_DOE}. Thus, a detector system must be developed to verify the polarization value at any given time. For polarized $^3\text{He}^{2+}$ ion beams elastic scattering processes can be used where the analyzing powers in the MeV range are known. In this case, a pre-acceleration step is required~\cite{Analyzing}. In this article, a new concept for a nuclear $^3\text{He}$ spin polarization detector based on the Lamb-shift polarimeter (LSP)~\cite{Lamb_pol} is introduced. The Lamb-shift polarimeter has been successfully implemented at different facilities, e.g.~for the polarized proton/deuteron source at TUNL~\cite{Source_TUNL}, the optically pumped polarized source at RHIC~\cite{History_LSP3} or the polarized $\text{H}^-$/$\text{D}^-$ source for the COSY storage ring~\cite{COSY_source, SPIN2025_Puetz}. Due to the very similar spin structure of the hydrogen atom and the $^{3}\text{He}^{+}$ ion the same selective quenching process can be applied. One of its advantages is that it works at low beam energies up to 100~keV which will fit to the beam energies of possible $^3\text{He}^{+(+)}$ ion sources itself. Of course, its data acquisition can be automated allowing for real-time analysis~\cite{SPIN2025_Puetz}.\\
In addition, such a polarimeter could also be used to detect and investigate the bound tritium decay, $T \rightarrow {^3He} + \bar{\nu_e}$. In this experiment, the LSP would not only serve to selectively detect the ejectiles  but could also simultaneously determine the nuclear spins of the $^3\text{He}$ nucleus, thereby enabling conclusions to be drawn about the chirality of the anti-electron neutrino analog to the so-called bound-beta decay of the free neutron~\cite{BoB}.
\newpage
\section{The Lamb-shift polarimeter}
\begin{figure}[ht]
\centering
\captionsetup{belowskip=-10pt} 
\includegraphics[scale=0.5, trim=50 0 50 0, clip]{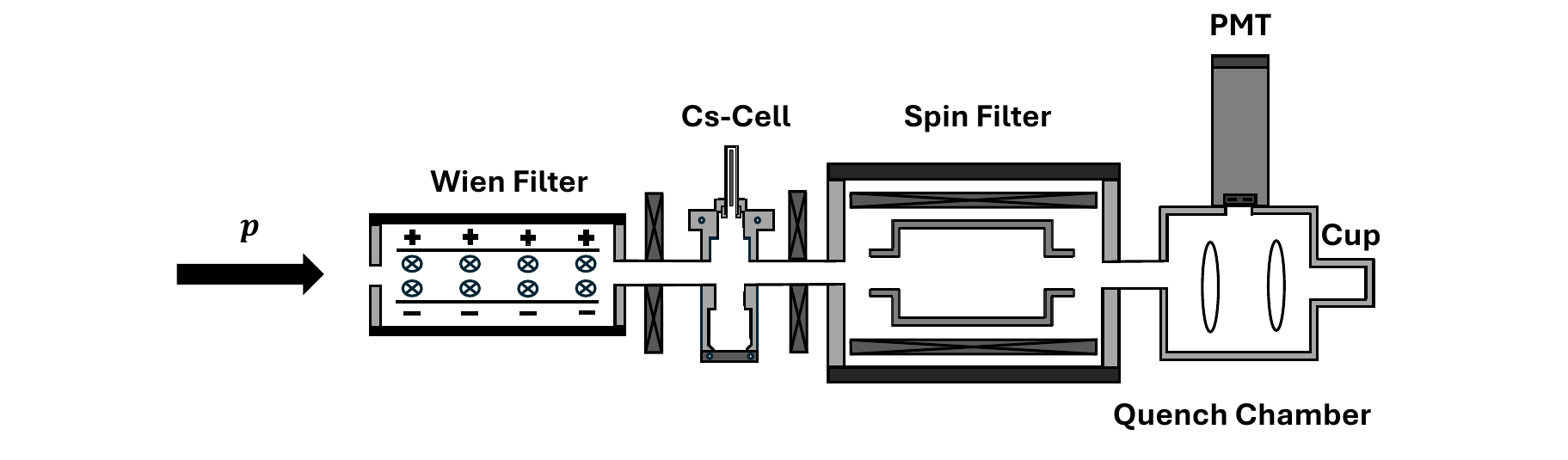}
\caption{\label{BILD LSP} The sketch shows a typical LSP setup consisting of a Wien filter, a cesium cell, a nuclear spin filter and a quenching region.}
\end{figure}

The classical LSP~\cite{Lamb_pol}, as illustrated in Fig.~\ref{BILD LSP}, consists of several different devices to measure the nuclear spin polarization of an incoming particle beam. The LSP can analyze all atoms, ions, or molecules composed of hydrogen or its isotopes at low kinetic energies~\cite{Lamb_pol,Ralf_Wien,Ralf_H_3_China}. The relative occupation numbers of different spin states are evaluated and then converted into a nuclear spin polarization value. Without loss of generality, the functionality of the LSP is shortly explained based on an incoming proton beam.
\FloatBarrier
\subsection{Wien Filter}
The first device, encountered by the particle beam is the Wien Filter. Its task is to ensure that only particles with a fixed velocity can pass through. Subsequently, it is capable of deflecting all charged particle species other than protons for a fixed beam energy.
In addition, the Wien filter is used to rotate the spin axis of the incoming particle beam, as the LSP is only sensitive to a longitudinal spin polarization along the beam axis. The magnetic field perpendicular to the beam direction induces a Larmor precession on the incoming magnetic moments/spins of the
particles, which is proportional to its magnetic flux $B_0$. So-called Wien filter curves, i.e.~the projection of the magnetic moments on the beam axis for different $B_0$, are visible in~\cite{Tarek_LHCb}. 
\subsection{Cesium cell}
Inside the cesium cell, the proton beam undergoes a charge exchange reaction to be transformed into hydrogen atoms. Thus, theoretical up to 30$\%$ of the hydrogen atoms end up in the excited metastable $2S_{\nicefrac{1}{2}}$ state, which is described in the following reaction~\cite{Cross_section_Cs,Recharge_H+_He+_in_Cs}
\begin{equation}
    \text{H}^++\text{Cs}\rightarrow \text{H}_{2S_{\nicefrac{1}{2}}}+\text{Cs}^+.
\end{equation}
Here, cesium is used for different reasons. First, it is an alkali metal, which means it has a low ionization energy. Second, its boiling point of approximately $941.4$~K is significantly lower than that of other alkali metals.\\
Similar reactions will appear for other ions like $H_2^+, H_3^+, H^-$ and even for fast hydrogen atoms within the keV energy range. When in a first step the electrons are stripped away, a proton is initially formed, which can then reabsorb an electron from the cesium to produce metastable hydrogen atoms. Although this reduces the efficiency of this transfer process depending on the particle and its energy, it is still high enough to yield a sufficient number of metastable atoms.\\ 
As long as these reactions take place in a strong magnetic field that decouples the nuclear and electron spins, the nuclear spin remains fully preserved during these reactions.
\subsection{Nuclear Spin Filter}
\begin{figure}[ht]
\centering
\captionsetup{belowskip=-10pt} 
\scalebox{0.8}[0.8]{\includegraphics{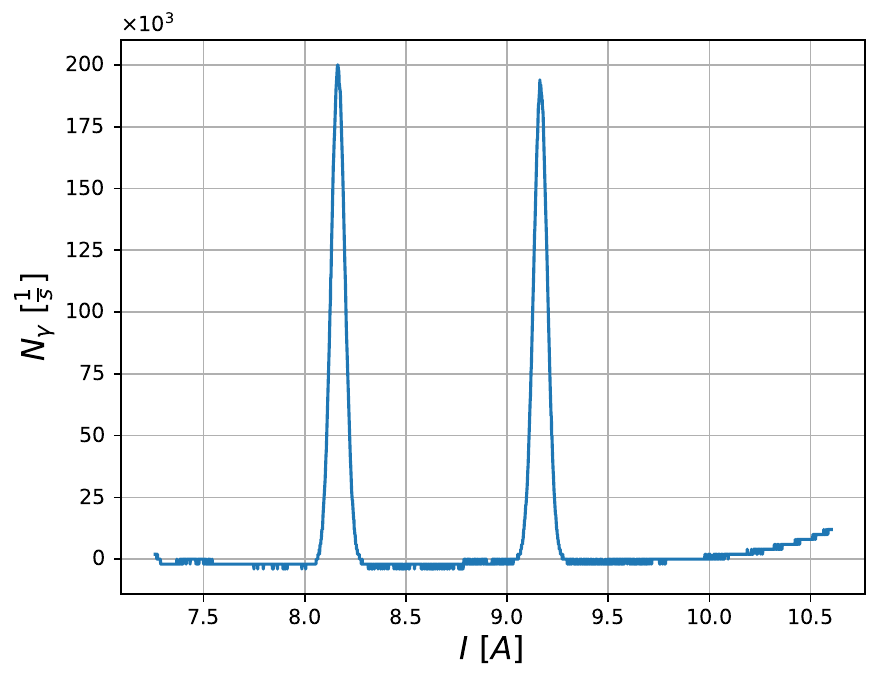}}
\caption{\label{ALL0016} The number of Ly-$\alpha$ photons, detected during the quenching process of metastable hydrogen atoms, is given as a function of the current $I$ inside the nuclear spin filter to produce the homogeneous magnetic field $B_z$ in beam direction. Thereby, the beam travels at an energy of 1~keV.}
\end{figure}

McKibben et al.~\cite{Los_Alamos,SF} provided the theoretical framework and the first realization of a nuclear spin filter. Its goal is to quench three of the four hyperfine states in the beam, while the remaining one is unaffected. Thereby, a three step process is applied using static electric and magnetic fields as well as a radio frequency. By scanning through the magnetic field area two conditions are found, where hydrogen atoms with different nuclear spin projection $m_I$ pass the spin filter. By comparing the number of metastable hydrogen atoms as function of the applied magnetic field results in the initial nuclear polarization value. Such a measurement is visible in Fig.~\ref{ALL0016}. 
\subsection{Quenching region}
The remaining metastable hydrogen atoms enter the quenching region, where strong electric fields couple them to the ground state. The emitted Ly-$\alpha$ photons are thereby detected by a photomultiplier.
\FloatBarrier

\section{Theoretical Framework}
\begin{figure}[ht]
\centering
\captionsetup{belowskip=-10pt} 
\scalebox{0.8}[0.8]{\includegraphics{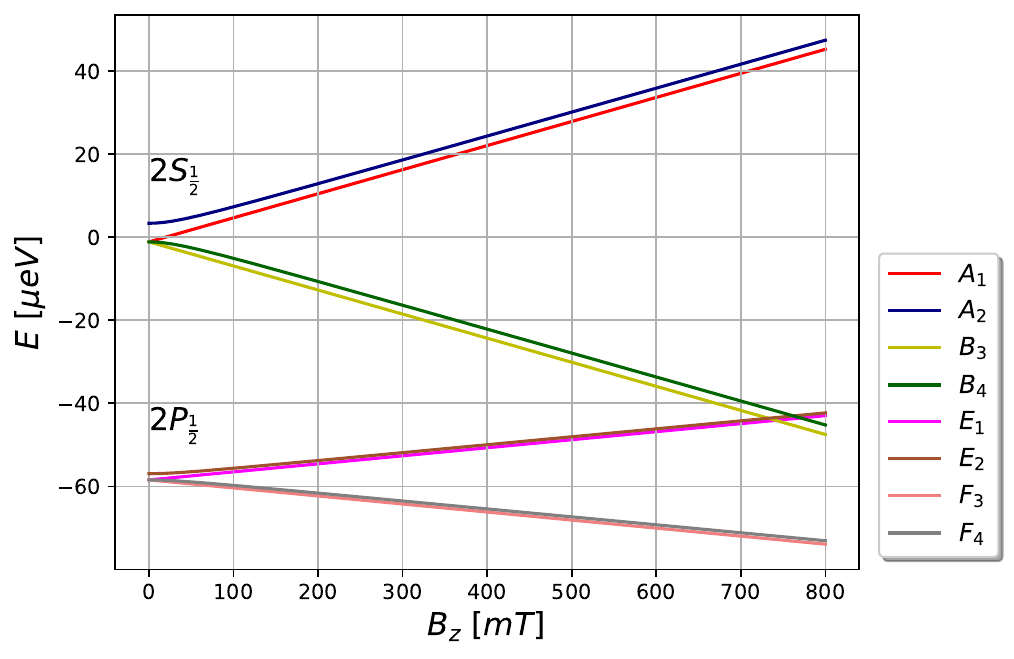}}
\caption{\label{BR_Helium} The $^3He^+$ eigenenergies for the Breit-Rabi Hamiltonian, introduced in Eq.~\eqref{BR-Ham}, are displayed as a function of the external magnetic field $\vec{B}=B_z\hat{e}_z$. The sets introduced here correspond to the metastable $2S_{\nicefrac{1}{2}}$ set and the short living $2P_{\nicefrac{1}{2}}$ set of the singly ionized $^3\text{He}^+$. Both are separated by the well-known Lamb-shift. The $y$-axis is scaled to show only the binding energy difference between the two sets of states.}
\end{figure}

The method~\cite{faatz2} presented in this work is based on the energy splitting in hydrogen-like systems and its interaction with external electromagnetic fields. Hy\-dro\-gen-like systems consist of a positively charged nucleus and one electron orbiting around it. 
The interaction of the electron spin $\vec{J}$ with the nuclear spin $\vec{I}$ is responsible for the hyperfine splitting. In addition, external magnetic fields $\vec{B}$ couple to any magnetic moment, which are produced by the total angular momentum of the electron $\vec{J}=\vec{S}\otimes\mathds{1}+\mathds{1}\otimes\vec{L}$, and by the nuclear spin $\vec{I}$. The corresponding Hamiltonian, which also includes the hyperfine splitting, is expressed as
\begin{equation}\label{BR-Ham}
H_{BR}=A\frac{\vec{I}\cdot\vec{J}}{\hbar^2}+\left(g_J\mu_B\frac{\vec{J}}{\hbar}-g_I\mu_k\frac{\vec{I}}{\hbar}\right)\cdot\vec{B}.
\end{equation}
The Bohr magneton is given by $\mu_B=\nicefrac{e\hbar}{2m_e}$ while the nuclear magneton is defined as $\mu_k=\nicefrac{e\hbar}{2m_p}$. Furthermore, $g_j$ is the Landé g-factor and $g_I$ is the nuclear g-factor. The remaining parameter $A$ serves as the hyperfine constant. Diagonalizing the Hamiltonian in case of an homogeneous external magnetic field $\vec{B}=B_z\hat{e}_z$ leads to the Breit-Rabi energies and their corresponding eigenstates. These are specifically given in~\ref{BR-States} and shown in Fig.~\ref{BR_Helium} with the additional Lamb-shift~\cite{Lamb_shift} for the excited $2S_{\nicefrac{1}{2}}$ and $2P_{\nicefrac{1}{2}}$ sets of singly ionized helium-3 ions. The corresponding values for the parameters are summarized in Tab.~\ref{Tabelle_Para_He}.
\begin{table}
\caption{\label{Tabelle_Para_He} The table summarizes the parameters for the singly ionized metastable $2S_{\nicefrac{1}{2}}$ and the $2P_{\nicefrac{1}{2}}$ $^3\text{He}^+$ ion states.}
\begin{tabularx}{\textwidth}{lp{1.5cm}Xp{3.35cm}p{1.5cm}X}
\toprule
&$g_j$\parnote{Introduced in~\ref{g_j}}&$g_I$&$A$~$\left[\text{MHz}\right]$&$\tau$&$\Delta E$~$\left[\text{MHz}\right]$\\
\midrule
$2S_{\nicefrac{1}{2}}$&$2.002$~\cite{He}&$-4.255$~\cite{He}&$-1083.355$~\cite{2SA_He_3,2SA_He_3_2}&$\frac{1}{525}$~s~\cite{Lifetime_Helium_3}&\\
$2P_{\nicefrac{1}{2}}$&$0.666$&$-4.255$~\cite{He}&$-360.801$\parnote{Utilized~\ref{A}}&$1.6$~ns~\cite{Lifetime2P}&$14041.474$~\cite{Lamb-shift_helium}\parnote{The Lamb-shift}\\
\bottomrule
\end{tabularx}
\parnotes
\end{table}
\FloatBarrier
Moreover, external electric fields $\vec{E}$ have an influence on the energy levels of hydrogen-like systems. The electric dipole interaction, also known as Stark effect, couples states with different orbital momenta to each other. The corresponding interaction is described by
\begin{equation}\label{Stark_Ham}
    H_{Stark}=e\vec{E}\cdot\vec{r},
\end{equation}
where $\vec{r}$ is the position operator of the electron, relative to the nucleus.\\ 
In addition to static fields, electromagnetic waves are also used in experimental setups. These waves are induced in cavities. A cylindrical cavity with height $d$ and base radius $R$ can have two different types of electromagnetic modes. In our case, the transversal magnetic modes (TM) are of great relevance. The corresponding fields in vacuum are represented in cylindrical coordinates as follows
\begin{equation}
    \begin{split}
        &E_r=-\frac{kx_{m,n}E_0}{2R\left(\mu_0\epsilon_0\omega^2-k^2\right)}\sin\left(kz\right)e^{-im\varphi}e^{-i\omega t}\left[J_{m-1}\left(\frac{x_{m,n}r}{R}\right)-J_{m+1}\left(\frac{x_{m,n}r}{R}\right)\right],\\
		&E_{\varphi}=\frac{im kE_0}{r\left(\mu_0\epsilon_0\omega^2-k^2\right)}\sin\left(kz\right)e^{-im\varphi}e^{-i\omega t}J_m\left(\frac{x_{m,n}r}{R}\right),\\
		&E_z=E_0J_{m}\left(\frac{x_{m,n}r}{R}\right)e^{-im\varphi}e^{-i\omega t}\cos\left(kz\right),\\
		&B_r=-\frac{\mu_0\epsilon_0\omega mE_0}{r\left(\mu_0\epsilon_0\omega^2-k^2\right)}\cos\left(kz\right)e^{-im\varphi}e^{-i\omega t}J_m\left(\frac{x_{m,n}r}{R}\right),\\
		&B_{\varphi}=\frac{i\mu_0\epsilon_0\omega x_{m,n}E_0}{2R\left(\mu_0\epsilon_0\omega^2-k^2\right)}\cos\left(kz\right)e^{-im\varphi}e^{-i\omega t}\left[J_{m-1}\left(\frac{x_{m,n}r}{R}\right)-J_{m+1}\left(\frac{x_{m,n}r}{R}\right)\right],\\
		&B_z=0.\\
    \end{split}
\end{equation}
The corresponding angular frequencies are stated as
\begin{equation}
    \omega_{m,n,p}=\frac{1}{\sqrt{\mu_0\epsilon_0}}\sqrt{\frac{x_{m,n}^2}{R^2}+\frac{p^2\pi^2}{d^2}},
\end{equation}
where $m,p\in\mathds{N}_0$ and $n\in\mathds{N}$. The n-th root $x_{m,n}$ of the Bessel functions of the first kind $J_m(x)$ plays an important role in obtaining the correct resonance frequency. Furthermore, the wave numbers $k$ are related to $p$ by
\begin{equation}
  k=\frac{p\pi}{d}.  
\end{equation}
The Hamiltonian for the electric components is modeled via Eq.~\eqref{Stark_Ham}, while the magnetic components of the electromagnetic waves follow the Breit-Rabi Hamiltonian (Eq.~\eqref{BR-Ham}) without the hyperfine splitting. 
So far, the theory was presented under the assumption that the cavity material is ideal, which allows only a single resonance frequency to enter at a time. However, by permitting the electromagnetic fields to penetrate the material, different modes around the resonance frequency, weighted by a Lorentz distribution
\begin{equation}
    f\left(\omega,\omega_{m,n,p},Q\right)=\frac{1}{\left(\omega-\omega_{m,n,p}\right)^2+\frac{\omega_{m,n,p}^2}{4Q^2}},
\end{equation}
may enter the cavity. The critical parameter here is the quality factor $Q$, which directly affects the full-width-half-maximum of the Lorentz curve. For this paper, the same ma\-the\-ma\-ti\-cal description as previously introduced in~\cite{FAATZ} is used.\\
Finally, the time evolution for the spin dynamics is described by the Lindblad master equation~\cite{Lindblad}
\begin{equation}
    \dot{\rho}\left(t\right)=-\frac{i}{\hbar}\left[H(t),\rho(t)\right]+\sum_{i=1}^{N}\gamma_i\left(\sigma_i\rho(t)\sigma^{\dagger}_i-\frac{1}{2}\left\{\sigma^{\dagger}_i\sigma_i,\rho(t)\right\}\right).
\end{equation}
The density operator $\rho(t)$ thereby describes an ensemble, and its diagonal elements give the probabilities of finding the single states, whereas $H(t)$ is the Hamiltonian of the system. Their commutator relation is also known under the name Liouville-von Neumann equation. The second part of the equation describes the dissipation of the population into the ground state, with $\gamma_i=\nicefrac{1}{\tau_i}$ being the damping factor, which is proportional to the inverse lifetime of a single excited state. Thereby, the different lifetimes of the two set of states, introduced in Tab.~\ref{Tabelle_Para_He}, play a major role in the functionality of the spin filter. Subsequently, the ladder operators $\sigma_i=\ket{g}\bra{i}$ connect these excited states $\ket{i}$ to the ground state $\ket{g}$. Last but not least, the anti-commutator is expressed by
\begin{equation}
    \left\{A,B\right\}=AB+BA.
\end{equation}

\section{LSP for $^3\text{He}^{+(+)}$ ions and atoms}
\begin{figure}[ht]
\centering
\captionsetup{belowskip=-10pt} 
\scalebox{0.5}[0.5]{\includegraphics{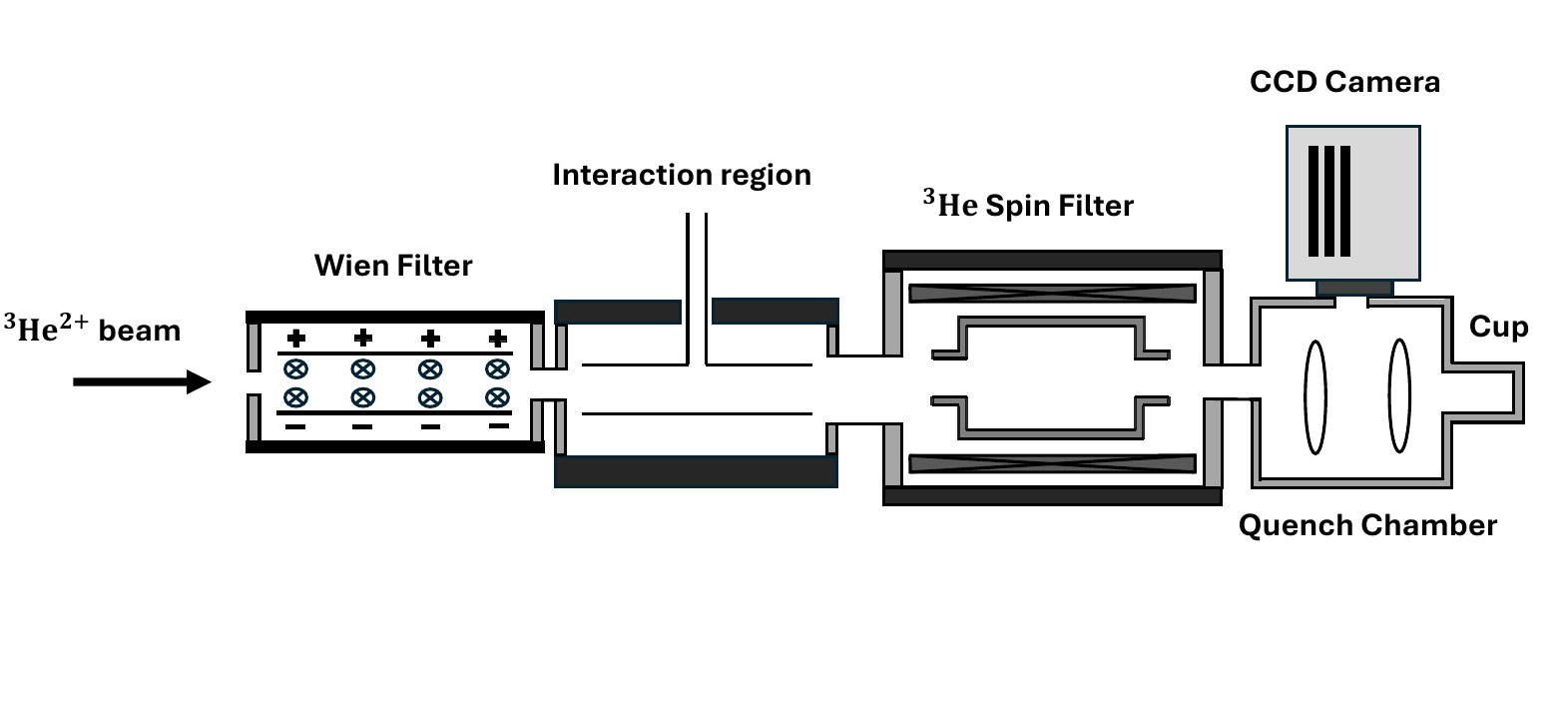}}
\caption{\label{LSP_He} The sketch shows a possible realization for a helium-3 LSP.}
\end{figure}

The LSP has existed for decades and has been successfully used in several different experiments, e.g.~\cite{LSP_Groningen,COSY,LSP_Russia,LSP_China}. This article has the intention to extend the use of the LSP beyond the current knowledge and to use it for $^3\text{He}^+$ ions. Besides the Wien Filter, the individual devices need to be adapted to be used for $^3\text{He}^+$  ions. A sketch of a possible final LSP set-up is shown in Fig.~\ref{LSP_He}.

\subsection{Production of metastable $^3\text{He}^+$ ion states}
The neutral helium atom carries two electrons. Consequently, there are two different types of ions, which are expected to be used by an LSP. The double ionized $^3\text{He}^{++}$ ion needs to recapture an electron to pass into the metastable $^3\text{He}^+$ state, while the singly ionized ground state needs to be excited into the metastable state via an inelastic scattering process. Similar to the measurement of proton polarization in the various forms of hydrogen (p, H, H$_2$, H$^-$), even $^3\text{He}$ and its ions can be converted into metastable $^3\text{He}^+$ ions for further use. \\
In 1974 Shah and Gilbody measured the efficiency of producing metastable $^3\text{He}^+$ ions for an incoming $^3\text{He}^{2+}$ beam at a beam energy of $10$ to $60$~keV by scattering on different gases~\cite{Reaction_He2+}. Later they published a follow up paper~\cite{Metastable_He_charge2} in which they extended the beam energy from $4$ to $434$~keV. This time, they measured the cross sections for creating metastable $^3\text{He}^+$ ions for an incoming $^3\text{He}^{2+}$ and $^3\text{He}^+$ ion beam on molecular and atomic hydrogen gas targets. These measurements show that metastable ions can be produced with an efficiency of about $13\%$ out of a $^3\text{He}^{2+}$ beam for both types of targets at a beam energy of approximately $40$~keV. Subsequently, the cross sections in the scattering process for an incoming singly ionized helium beam is nearly two orders of magnitude smaller and beam energies up to $100$~keV are necessary. Nevertheless, for both species it is possible to produce metastable $^3\text{He}^{+*}$ ions and further experimental measurements for different gas targets can be performed to find the most suitable option. \\
$^3\text{He}$ atoms in a slow beam at thermal energies~\cite{ABS_He} can be ionized in a strong magnetic field to preserve the nuclear polarization by decoupling it from the electron spin and accelerate the produced $^3\text{He}^{+(+)}$ ions up to the necessary beam energies described before.

\subsection{$^3\text{He}$ spin filter}
The spin filter is the main device, that separates the single Beit-Rabi states and makes the determination of the nuclear spin polarization possible.

\subsubsection{Spin dynamics}
\begin{figure}[ht]
\centering
\captionsetup{belowskip=-10pt} 
\scalebox{0.8}[0.8]{\includegraphics{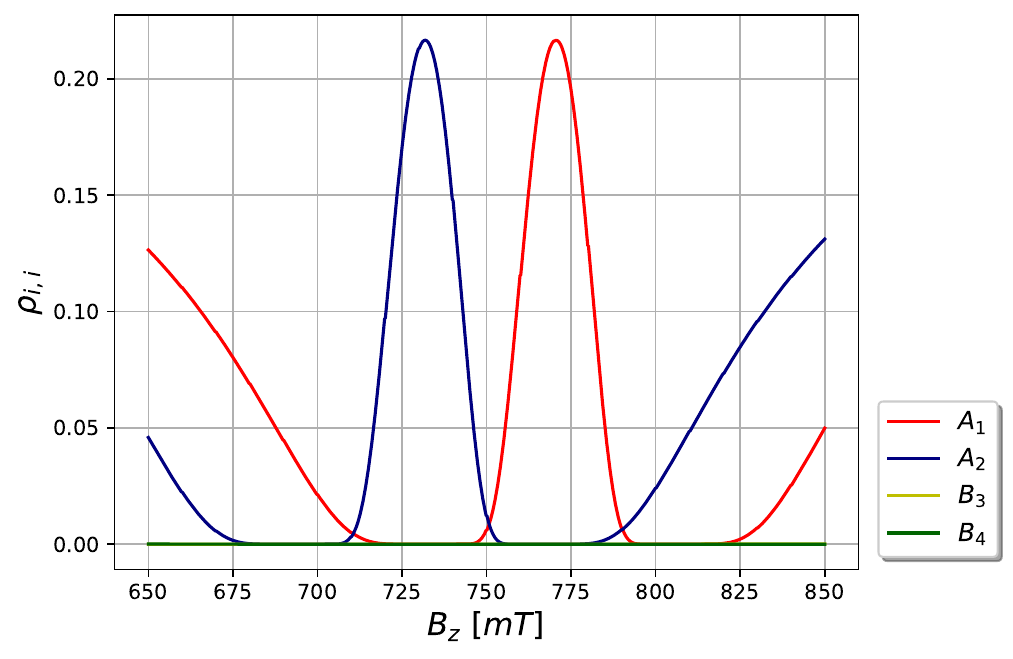}}
\caption{\label{Spin_dynamics_1} The probabilities of the metastable singly ionized $^3\text{He}^+$ states are plotted as function of the external magnetic field $B_z$ for an unpolarized initial beam. Thereby, the ions travel with an energy of 1.5~keV and the parameters for the simulation are obtainable in Tab.~\ref{Tabelle_Para_Sim}. For completeness, the $B_3$ and $B_4$ states are equally zero.}
\end{figure}

\begin{figure}[ht]
\centering
\captionsetup{belowskip=-10pt} 
\scalebox{0.8}[0.8]{\includegraphics{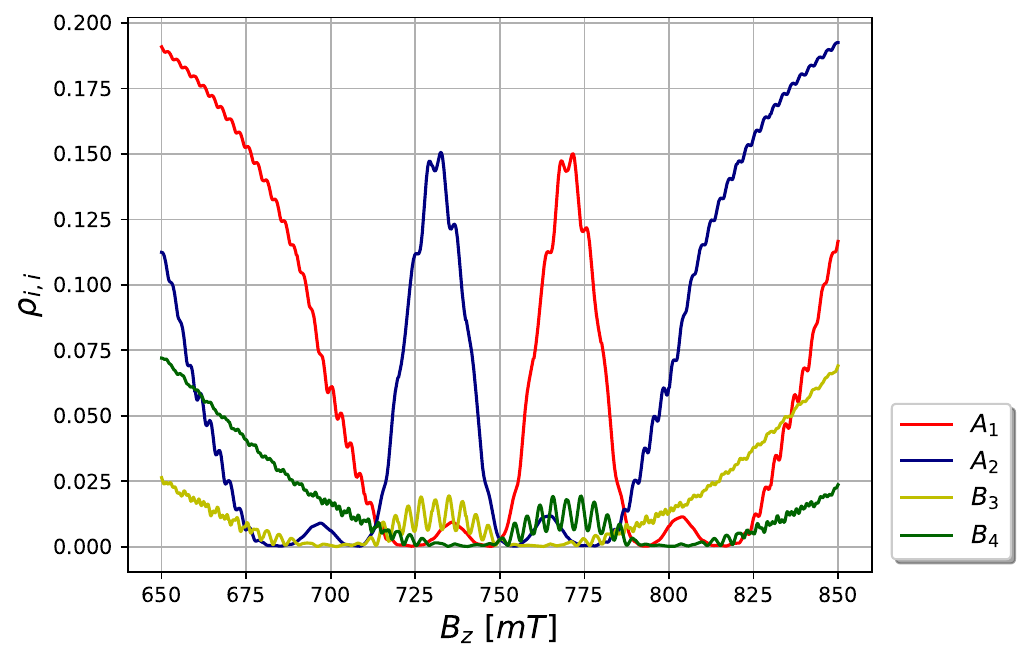}}
\caption{\label{Spin_dynamics_2} The probabilities of the metastable $^3\text{He}^+$ states are plotted as function of the external magnetic field $B_z$ for an unpolarized initial beam. Thereby, the ions travel with an energy of 100~keV and the parameters for the simulation are obtainable in Tab.~\ref{Tabelle_Para_Sim}.}
\end{figure}

By utilizing external electromagnetic fields it is possible to manipulate the metastable $^3\text{He}^{+}$ ion beam such that only ions in a dedicated hyperfine state remain in the beam. Thus, the selective quenching process consists of three steps. First, a homogeneous external magnetic field $\vec{B}=B_z\hat{e}_z$ along the beam axis is applied to split the eigenergies of the hyperfine states and to enter the Paschen-Back regime, as visible in Fig.~\ref{BR_Helium}. At magnetic field values from $700$~mT up to $790$~mT the eigenenergies of the metastable $^3\text{He}^+$ ion states $B$ cross those of the short-lived $E$ states of the $2P_{\nicefrac{1}{2}}$ set. As the $2S_{\nicefrac{1}{2}}$ and the $2P_{\nicefrac{1}{2}}$ sets have a parity difference, a static electric field $\vec{E}=E_x\hat{e}_x$, perpendicular to the beam axis, is applied to couple the $B$ states at the crossing points to the $E$ states. Due to this coupling, the lifetimes $\tau_{B}$ of the $B$ states are highly reduced, causing them to decay into the ground state. In the last step, a radio frequency is employed into a cylindrical cavity. Its frequency of $f=21.0636$~GHz fits to the energy gap between the crossing point area and the $A$ states. Thus, the corresponding $A$ state is repopulated by the short increase in the population of the $E$ states if the magnetic flux is applied at the matching crossing point. Meanwhile, the second $A$ state decays slowly. By ramping through the magnetic field, the two distinct $A$ peaks appear at the respective value for the energy crossing points. The peak heights of the two states, carrying different nuclear spin projections $m_I$, are compared to identify the initial nuclear spin polarization of the beam.
\newline
In comparison to the hydrogen atom, the energy gap caused by the Lamb-shift is significantly larger due to the nuclear charge number of $Z=2$. This explains also that any other set of states can be neglected for the calculations. The closest set is the $2P_{\nicefrac{3}{2}}$, which has an energy gap of approximately $526~\mu\text{eV}$ at 800~mT towards the nearest metastable $A$ state.
\newline
Subsequently, we suggest to use the $\text{TM}_{0,1,10}$ mode to produce the large frequency of $f=21.0636$~GHz while having a realistic size for the cavity and for the beam line. With this mode, a base radius of $R=4$~cm and a length of $d\approx7.183$~cm can be reached, which provides enough space for a beam line. Corresponding simulations for the spin dynamics in case of an initially unpolarized beam are depicted in Fig.~\ref{Spin_dynamics_1} and Fig.~\ref{Spin_dynamics_2}. The necessary parameters to perform the simulations are listed in Tab.~\ref{Tabelle_Para_He} and Tab.~\ref{Tabelle_Para_Sim}. Even for large beam energies around $E_{Kin}=100$~keV, the separation between the two $A$ states is still possible, as it is visible in Fig.~\ref{Spin_dynamics_2}. As a consequence, some intensity is lost in comparison to the slower beam simulated in Fig.~\ref{Spin_dynamics_1}. In addition, we applied a switch of polarity in the static electric field to ensure the beam is not defocused. This does not significantly affect the peak structure of the $A$ states. This part is investigated in more detail in Sec.~\ref{Beam_trac}\\ 
Moreover, to achieve the large magnetic fields we would advise to use superconducting magnets. The cavity would also take advantage of the low temperature, as $Q$-values around $16000$ are needed. At low temperatures, materials like silver exhibit a significant decrease in resistivity, reaching values as low as $\rho=10^{-11}~\Omega\text{m}$~\cite{Ag_resistivity}. 
This affects the $Q$-value drastically, as its dependence based on the material is stated as~\cite{Q_value_metall}
\begin{equation}
    Q=0.38\,\frac{\lambda_0}{\delta}\frac{1}{1+\frac{R}{d}},
\end{equation}
where $\lambda_0$ denotes the resonance wavelength of the electromagnetic wave coupled into the cavity, and $\delta$ the skin depth
\begin{equation}
    \delta=\sqrt{\frac{2\rho}{\omega_0\mu}}\sqrt{\sqrt{1+\left(\rho\omega_0\epsilon\right)^2}+\rho\omega_0\epsilon}.
\end{equation}
\begin{table}
\caption{\label{Tabelle_Para_Sim} The table provides the reader with the parameters to perform the simulations leading to Fig.~\ref{Spin_dynamics_1} and Fig.~\ref{Spin_dynamics_2}. In addition, the velocity is related to the beam energy by $E_{kin}=\nicefrac{m_{^3\text{He}}v^2}{2}$.}
\begin{tabularx}{\textwidth}{XXXXXXX}
\toprule
Figure&$E_x$~$\left[\frac{\text{kV}}{\text{m}}\right]$\parnote{Electric field switches polarity in the middle of the cavity}&$E_0$~$\left[\frac{\text{kV}}{\text{m}}\right]$&$v$~$\left[\frac{\text{m}}{\text{s}}\right]$&$E_{kin} \left[\text{keV}\right]
$ &$r$~$\left[\text{m}\right]$&$Q$\\
\midrule
Fig.~\ref{Spin_dynamics_1}&$50$&$30$&$309500$&1.5&0&$16000$\\
Fig.~\ref{Spin_dynamics_2}&$50$&$30$&$2526451$&100&0&$16000$\\
Fig.~\ref{Spin_dynamics_3}&$50$&$30$&$1786470$&50&0.01&$16000$\\
\bottomrule
\end{tabularx}
\parnotes
\end{table}
\FloatBarrier
\FloatBarrier
\begin{figure}[ht]
\centering
\captionsetup{belowskip=-10pt} 
\scalebox{0.8}[0.8]{\includegraphics{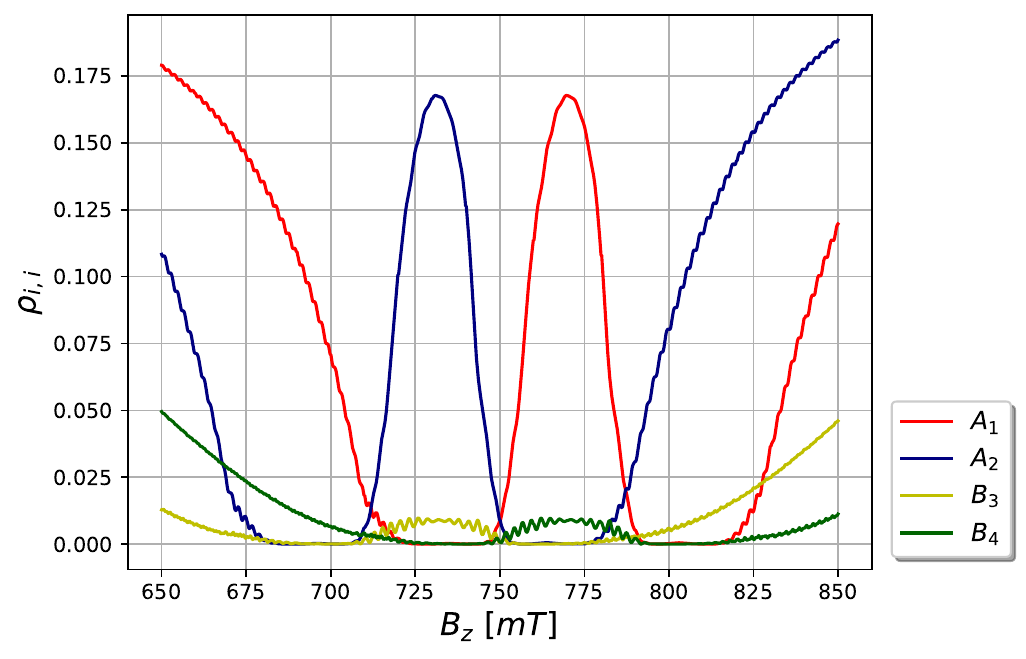}}
\caption{\label{Spin_dynamics_3} The probabilities of the metastable $^3\text{He}^+$ states are plotted as function of the external magnetic field $B_z$ for an unpolarized initial beam. Thereby, the ions travel with an energy of 50~keV at a fixed radius of 1~cm and the parameters for the simulation are obtainable in Tab.~\ref{Tabelle_Para_Sim}.}
\end{figure}
\FloatBarrier

\subsubsection{Beam trajectories}\label{Beam_trac}
In contradiction to the hydrogen atoms in a classical spinfilter, the external electric and magnetic fields strongly influence a charged beam of $^3$He$^+$ ions due to the Lorentz force. Since the direction of the magnetic field along the axis is parallel to the ions’ trajectory, its effect on deflection is rather small and can be ignored. Depending on the ions’ velocity, there are various ways to handle the deflection induced by the vertical electric fields. For example, the electric field can be split into two separated parts with opposite field directions to compensate the deflection. Nevertheless, at very high beam energies of nearly 100 keV, the deflection at the end of the spin filter is very small anyway ($< 1$~mm) and can be compensated for in the quench chamber without major effort. At lower energies, this deflection can be even used to separate the $^3$He$^+$ ions from neutral atoms or doubly charged $^3$He$^{++}$ ions using slit apertures, thereby minimizing the background in the measured spectra.

\subsection{$^3\text{He}^+$ quenching region}
\begin{figure}[ht]
\centering
\captionsetup{belowskip=-10pt} 
\scalebox{0.4}[0.4]{\includegraphics{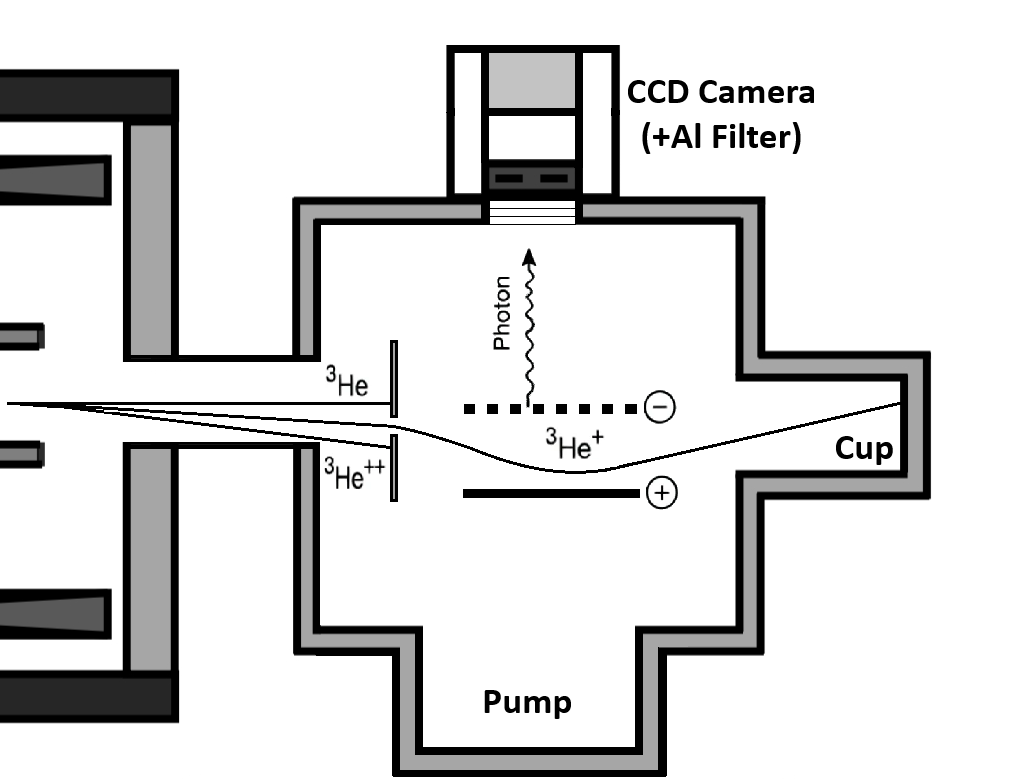}}
\caption{\label{CCD_camera_He} 
The sketch shows a possible arrangement for the electric quenching field. Therefore, high voltage can be applied between an electrode and the grid. The tasks of this electric potential are to de-excite the metastable $^3\text{He}^+$ ions and to refocus the beam towards the cup. The induced Lyman-$\alpha$ photons can enter a CCD camera through the grid.
}
\end{figure}

In the quenching region, the number of metastable $^3\text{He}^+$ ions is detected. In case of hydrogen, the metastable atoms are quenched by strong electric fields, causing the emitted Ly-$\alpha$ photons to trigger a cascade of electrons on the photo-cathode of a photomultiplier. These photons have roughly an energy of $10.2$~eV and a corresponding wavelength of $121.6$~nm, which requires special photomultiplier (Photocathode: KBr, Entrance window: MgF$_2$) to reduce the background radiation. Contrarily, due to the larger charge number $Z=2$ the photons emitted in the quenching process of metastable $^3\text{He}^+$ ions, due to stronger electric fields, have an energy of $40.8$~eV and a corresponding wavelength of $30.4$~nm. This photon energy cannot be detected by classical photomultiplier and,  consequently, they need to be replaced. Micro-channel plates could be used, but instead, we suggest using a CCD camera, which has a high quantum efficiency up to $75\%$ around $40$~eV~\cite{CCD}. In Fig.~\ref{CCD_camera_He} a sketch for parts of a possible quenching field arrangement is shown. A grid can help to identify the positional de-excitation of the metastable $^3\text{He}^+$ ions, while the applied voltage helps to refocus the $^3\text{He}^+$ ion beam towards the cup, making a polarity change in the nuclear spin filter redundant. A Beryllium/Magnesium coating on the lower electrode would reflect about 56$\%$~\cite{reflection} of the incoming photons in this direction and, therefore, increase the signal amplitude. \\
Like discussed before, an additional aperture behind the spinfilter will be able to stop the neutral atoms and the $^3$He$^{2+}$ ions to suppress the background in the measured signals. 
To reduce the background even further, a thin aluminum foil ($\sim 1~\mu m$) can be placed in front of the camera. Its transmission probabilities for these photons are just about 5 $\%$, but it will cut off photon energies below 20 eV and above 70~eV~\cite{HENKE_Al,CXRO}. Thus, the corresponding background reduction might be worth to accept the intensity losses of the photon signal.
\FloatBarrier

\newpage
\section{Conclusion and Outlook}
Based on the LSP for hydrogen, we introduced a modified version to validate the nuclear polarization of $^3\text{He}^{+(+)}$ ions and $^3\text{He}$ atoms. The advantage is, that at low beam energies up to $100$~keV, a quick analysis with small uncertainties on the $\%$ level can be performed. This would prevent the need to pre-accelerate the beam for nuclear scattering polarimeters to identify the nuclear polarization value. In addition, it might serve as tool to verify the nuclear polarization of planned internal $^3\text{He}$ gas targets as proposed in~\cite{ABS_He}. \\
Another possible application of such a Lamb-shift polarimeter could be the direct detection of $^3\text{He}$ atoms resulting from the bound-beta decay of tritium \cite{tritium},\newline 
$T \rightarrow {^3\text{He} + \bar{\nu _e}}$, similar to the bound beta decay of the neutron~\cite{BoB}. In this decay, the $^3\text{He}$ atoms have a kinetic energy of 62~meV and can be separated from the $^3\text{He}^+$ ions by deflecting electric and magnetic fields. Diffusing $^3\text{He}$ atoms from recombined ions in the residual gas might be separated by a chopper between the tritium source and the polarimeter. The fixed velocity of about 2000~m/s of the $^3\text{He}$ atoms from the the bound-beta decay of tritium will determine the time-of-flight of these atoms from the chopper to the ionizer. Afterwards, the $^3\text{He}$ atoms could be ionized into $^3\text{He}^{+(+)}$ ions in a strong magnetic field, accelerated to the necessary beam energies and again excited into the metastable state $^3\text{He}^{+*}$ by inelastic collisions with Argon. It may even be possible to directly induce the atoms to enter the metastable state during ionization with a high cross section. 
After the individual hyperfine-structure substates have been filtered by the spin filter, the detection of Lyman-$\alpha$ photons would not only prove the existence of this decay, but would also provide information about the polarization of the nucleus and, consequently, about the chirality of the anti-electron neutrino $\bar{\nu _e}$.\\
Nevertheless, the realization of the nuclear spin filter, as well as optimizing the process for producing metastable $^3\text{He}^{+*}$ ions and to detect their quenching light, need to be experimentally addressed.

\section*{Acknowledgments}\label{sec.Acknowledgments}
We wish to thank Prof.~J.~Pretz and Prof.~C.~Hanhart for their supervision and help to realize this paper. In addition, C. Kannis acknowledges funding from the Deutsche Forschungsgemeinschaft (DFG, German Research Foundation) – 533904660.

\newpage
\appendix
\section{Breit-Rabi solutions for singly ionized $^3\text{He}^+$ ions}\label{BR-States}
In this part, the eigensolution for Eq.~\eqref{BR-Ham} is introduced for the metastable $2S_{\nicefrac{1}{2}}$ and the $2P_{\nicefrac{1}{2}}$ sets. Additionally, the Lamb-shift is included to the hyperfine splitting in form of the experimental value $\Delta E_{Lamb}$~\cite{Lamb-shift_helium}. The $2S_{\nicefrac{1}{2}}$ set is expressed in the $\Ket{F,m_F,j,l}$ basis
\begin{equation}
    \begin{split}
    &\Ket{A_1}=\Ket{1,1,\nicefrac{1}{2},0},\\
    &E_{A_{1}}=\frac{A_{2S_{\nicefrac{1}{2}}}}{4}+\frac{1}{2}\left(g_s\mu_B-g_I\mu_k\right)B_z,\\
    &\Ket{A_2}=\frac{1}{\sqrt{1+X^2(B_z)}}\left(\Ket{0,0,\nicefrac{1}{2},0}+X(B_z)\Ket{1,0,\nicefrac{1}{2},0}\right),\\
    &E_{A_{2}}=-\frac{A_{2S_{\nicefrac{1}{2}}}}{4}+\frac{1}{2}\sqrt{A_{2S_{\nicefrac{1}{2}}}^2+\left(g_s\mu_B+g_I\mu_k\right)^2B_z^2},\\
    &\Ket{B_3}=\Ket{1,-1,\nicefrac{1}{2},0},\\
    &E_{B_{3}}=\frac{A_{2S_{\nicefrac{1}{2}}}}{4}-\frac{1}{2}\left(g_s\mu_B-g_I\mu_k\right)B_z,\\
    &\Ket{B_4}=\frac{1}{\sqrt{1+\Omega^2(B_z)}}\left(\Omega(B_z)\Ket{0,0,\nicefrac{1}{2},0}+\Ket{1,0,\nicefrac{1}{2},0}\right),\\
    &E_{B_{4}}=-\frac{A_{2S_{\nicefrac{1}{2}}}}{4}-\frac{1}{2}\sqrt{A_{2S_{\nicefrac{1}{2}}}^2+\left(g_s\mu_B+g_I\mu_k\right)^2B_z^2},\\
    &\text{with}\quad X\left(B_z\right)=\frac{\left(g_s\mu_B+g_I\mu_k\right)B_z}{-A_{2S_{\nicefrac{1}{2}}}+\sqrt{A_{2S_{\nicefrac{1}{2}}}^2+\left(g_s\mu_B+g_I\mu_k\right)^2B_z^2}}\\
    &\text{and}\quad\Omega\left(B_z\right)=\frac{-A_{2S_{\nicefrac{1}{2}}}-\sqrt{A_{2S_{\nicefrac{1}{2}}}^2+\left(g_s\mu_B+g_I\mu_k\right)^2B_z^2}}{\left(g_s\mu_B+g_I\mu_k\right)B_z}.\\
    \end{split}
\end{equation}
Next follows the $2P_{\nicefrac{1}{2}}$ set
\begin{equation}
\begin{split}
    &\Ket{E_1}=\Ket{1,1,\nicefrac{1}{2},1},\\
    &E_{E_{1}}=-\Delta E_{Lamb}+\frac{A_{2P_{\nicefrac{1}{2}}}}{4}+\frac{1}{2}\left(g_j\mu_B-g_I\mu_k\right)B_z,\\
    &\Ket{E_2}=\frac{1}{\sqrt{1+Y^2(B_z)}}\left(\Ket{0,0,\nicefrac{1}{2},1}+Y(B_z)\Ket{1,0,\nicefrac{1}{2},1}\right),\\
    &E_{E_{2}}=-\Delta E_{Lamb}-\frac{A_{2P_{\nicefrac{1}{2}}}}{4}+\frac{1}{2}\sqrt{A_{2P_{\nicefrac{1}{2}}}^2+\left(g_j\mu_B+g_I\mu_k\right)^2B_z^2},\\
    &\Ket{F_3}=\Ket{1,-1,\nicefrac{1}{2},1},\\
    &E_{F_{3}}=-\Delta E_{Lamb}+\frac{A_{2P_{\nicefrac{1}{2}}}}{4}-\frac{1}{2}\left(g_j\mu_B-g_I\mu_k\right)B_z,\\
    &\Ket{F_4}=\frac{1}{\sqrt{1+Z^2(B_z)}}\left(Z(B_z)\Ket{0,0,\nicefrac{1}{2},1}+\Ket{1,0,\nicefrac{1}{2},1}\right),\\
    &E_{F_{4}}=-\Delta E_{Lamb}-\frac{A_{2P_{\nicefrac{1}{2}}}}{4}-\frac{1}{2}\sqrt{A_{2P_{\nicefrac{1}{2}}}^2+\left(g_j\mu_B+g_I\mu_k\right)^2B_z^2},\\
    &\text{with}\quad Y\left(B_z\right)=\frac{\left(g_j\mu_B+g_I\mu_k\right)B_z}{-A_{2P_{\nicefrac{1}{2}}}+\sqrt{A_{2P_{\nicefrac{1}{2}}}^2+\left(g_j\mu_B+g_I\mu_k\right)^2B_z^2}}\\
    &\text{and}\quad Z\left(B_z\right)=\frac{-A_{2P_{\nicefrac{1}{2}}}-\sqrt{A_{2P_{\nicefrac{1}{2}}}^2+\left(g_j\mu_B+g_I\mu_k\right)^2B_z^2}}{\left(g_j\mu_B+g_I\mu_k\right)B_z}.\\
\end{split}
\end{equation}

\section{Landé g-factor}\label{g_j}
In this section, the Landé g-factor $g_j$ is introduced. It is a combination of the orbital g-factor $g_l=1$~\cite{g_s_alt} and the spin g-factor $g_s\approx2.002$~\cite{g_s_neu}. Therefore, the single magnetic moments add up to a total magnetic moment, which is only possible if $g_j$ takes the following form  
\begin{eqnarray}
    \scalebox{1.2}{$g_j=\frac{1}{2}\frac{g_s\left[j\left(j+1\right)+s\left(s+1\right)-l\left(l+1\right)\right]+g_l\left[j\left(j+1\right)+l\left(l+1\right)-s\left(s+1\right)\right]}{j\left(j+1\right)}$.}
\end{eqnarray}
\section{Hyperfine constant}\label{A}
Based on~\cite{Bethe}, the Hamiltonian for the hyperfine splitting is introduced as
\begin{equation}
    H_{Hyp}=A\frac{\vec{I}\cdot\vec{J}}{\hbar^2}.
\end{equation}
This result is in the limit of transitions in the same set of states meaning the principal quantum number $n$ and orbital quantum number $l$ are fixed. In the derivation, the hyperfine constant $A$ is determined for $l=0$ as
\begin{equation}
    A=\frac{e^2g_I\hbar^2}{3\varepsilon_0c^2m_em_p}\frac{1}{4\pi}|R_{n,0}(r=0)|^2,
\end{equation}
and in the case of $l\neq0$ as
\begin{equation}
  A=\frac{e^2\hbar^2g_I}{2m_em_p4\pi\varepsilon_0c^2j(j+1)a_0^3n^3\left(l+\nicefrac{1}{2}\right)}.  
\end{equation}
\newpage
\bibliography{apssamp}

\end{document}